# The Second Amendment and Cyber Weapons

The Constitutional Relevance of Digital Gun Rights

Jan Kallberg
Army Cyber Institute at West Point (ACI)
and the Department of Social Sciences
United States Military Academy
West Point, NY 10996
jan.kallberg@usma.edu

*Abstract*— *In the future, the United States government can seek to limit the ownership and usage of cyber weapons. The question is whether the Second Amendment to the United States Constitution gives a right to bear and own military-grade cyber weapons, and if so, under which conditions. The framers of the Bill of Rights, ratified in 1791, did not limit the right to bear arms to defined weapons such as long rifles and pistols, but instead chose the broader word* arms. *The United States Supreme Court, in the case District of Columbia v. Heller, upheld a demarcation between "dangerous and unusual" weapons that are not permissible to own and weapons protected by the Second Amendment. Cyber weapons take the form of dual-use software, often shared and globally distributed, that in most cases can be weaponized for harmful purposes. In recent years, major corporations have sought to "hack back," and if the hack back is authorized, the question becomes whether corporations have digital gun rights. Even if corporations are considered US persons, they do not automatically obtain digital gun rights based on the Second Amendment. This article discusses the core constitutional challenges for the United States government in prohibiting individual ownership of cyber weapons and the rationale for why corporations are in a weaker position regarding ownership of cyber arms. The argument brought forward is that individuals can claim Second Amendment protection of their right to own military-grade software tools, but corporations must meet additional criteria to do so.*

## I. INTRODUCTION

If the US government pursued the regulation of cyber weapons and the prohibition of private ownership of these arms, the question becomes if and how the right to bear arms would protect ownership of cyber weapons from governmental intervention. Another question is whether the Second Amendment to the US Constitution provides individuals and corporations with equal protection regarding the right to own advanced cyber weapons. In most cases, public debate over the Second Amendment is ideological, but there also is legal doctrine given by the US Supreme Court in the verdict of District of Columbia v. Heller that culminates the aggregation of precedent covering more than two centuries.

The Second Amendment to the US Constitution, ratified 1791, was not only a product of the American Revolution and recognition of the state militias' contribution to the war effort, but also was an idea nurtured by early influencers of the Founding Fathers [1],[2]. A hundred years before the American Revolution, Hobbes declared that the right to bear arms was a response to the lingering chaos of human conflict and without access to weapons, society would fall into a state of entropy. Montesquieu considered armed citizens a counter-balance to tyranny and the abuse of power. During the Virginia Convention in June 1776, when Jefferson and Mason worked to word the new constitution for the State of Virginia, they wrote that "no freeman should be debarred the use of arms." This phrase did not appear in the final bill [3].

The Second Amendment to the US Constitution reads as follows: "A well-regulated militia, being necessary to the security of a free state, the right of the people to keep and bear arms, shall not be infringed." The authors of the Bill of Rights and Second Amendment could have listed the weapons of their era, but instead used the general term, *arms*. Restrictions on the weapons that citizens have the right to bear have been implemented over time through litigation and guidance from the US Supreme Court and precedents put forth through the legal system.

## II. DANGEROUS AND UNUSUAL WEAPONS

In the most recent legal challenge to Second Amendment doctrine, addressed by the US Supreme Court, a central legal question was what types of weapons are protected by the Second Amendment. If a weapon is "dangerous and unusual" under the US Supreme Court's interpretation and verdict, then ownership of such a weapon is not protected by the Second Amendment to the US Constitution. If a weapon is deemed "dangerous and unusual," then the government is not hindered by the Second Amendment to forbid ownership and usage.

In their District of Columbia v. Heller verdict, the US Supreme Court majority wrote [4]:

"We also recognize another important limitation on the right to keep and carry arms. Miller said, as we have explained, that the sorts of weapons protected were those 'in common use at the time.' 307 U. S., at 179. We think that limitation is fairly supported by the historical tradition of prohibiting the carrying of "dangerous and unusual weapons." See 4 Blackstone 148–149 (1769); 3 B. Wilson, Works of the Honourable James Wilson 79 (1804); J. Dunlap, The New-York Justice 8 (1815); C. Humphreys, A Compendium of the Common Law in Force in Kentucky 482 (1822); 1 W. Russell, A Treatise on Crimes and Indictable Misdemeanors 271–272 (1831); H. Stephen, Summary of the Criminal Law 48 (1840); E. Lewis, An Abridgment of the Criminal Law of the United States 64 (1847); F. Wharton, A Treatise on the Criminal Law of the United States 726 (1852). See also State v. Langford, 10 N. C. 381, 383–384 (1824); O'Neill v. State, 16Ala. 65, 67 (1849); English v. State, 35Tex. 473, 476 (1871); State v. Lanier, 71 N. C. 288, 289 (1874)."

The US Supreme Court majority refer to cases that define unusual and dangerous weapons, and also to behaviors that could endanger others. In State v. Lanier (1874), a drunk man rides at a canter pace at midnight through a courthouse yard, which is considered an endangerment. The verdict of the North Carolina Supreme Court rejected the prosecution's premise. The North Carolina Supreme Court based its decision on the conditions surrounding the event: "We conceive that the riding through a courthouse or a street at 12 o'clock at night, when no one is present, is a very different thing from riding through at 12 o'clock in the day, when the courthouse or street is full of people." The fact that the man rode at a canter pace through town at midnight was not a public endangerment and did not break the peace.

In English v. State, protected arms are defined as the arms of a militia, which are military-grade arms.

### III. Dangerous Cyberweapons

The US Supreme Court, with supporting legal precedence, has declared that only weapons that are not "dangerous and unusual" are permissible.

The next question is what would constitute a dangerous and unusual cyber weapon that would preclude ownership—including placement on computers—by citizens. First, the existence and effect of the weapon must be known to the government and the lawmaker so it could be restricted. A militiaman is armed with his personal armament: rifle, bayonet, and sidearm. In the nondigital world, as an example, claymore mines and hand grenades are not permissible weapons for the citizenry, but are considered "dangerous and unusual" and therefore restricted to government use. The damage and hazards surrounding claymore mines and hand grenades are known to the government and the lawmaker. The tripwire will detonate the claymore mine, and the hand grenade is thrown without a clear understanding of the final impact and effect. The mine is autonomous in its execution, and the grenade is not sufficiently accurate, with a possibility of damages beyond the intended scope of the defender. These effects are known to the government. In cyber, this decision path is different.

The lawmaker needs to be aware of cyberweapons and the means of restricting them. Otherwise, the result would be a ban on thought and innovation; the law would then arbitrarily punish what in retrospect is considered "dangerous and unusual cyberweapons" through law that is ex-post-facto legislation.

### IV. Definition of Cyber Arms

There no precise definition of cyber arms or the anticipated capabilities and effects of their utilization. One reason is that the very nature of software creates multiple purposes for utilization.

The US Department of Defense has no codified, uniform definition, but refers to cyber arms as a "capability:" "a device, computer program, or technique, including any combination of software, firmware, or hardware, designed to create an effect in or through cyberspace" [5].

The common definition of cyber arms is broad and comprehensive. One definition, by the North Atlantic Treaty Organization (NATO) Cooperative Cyber Defence Centre of Excellence (CCDOE), is [6] "…software, firmware or hardware designed or applied to cause damage through the cyber domain."

This definition is legally problematic because of the assumption that the intended usage of software, firmware, or hardware would alter the status of a nonweapon to a weapon. This contradicts elementary principles in the rule of law, where legal rules altered after the fact are referred to as ex-post facto laws. The cyber community have yet not presented a commonly accepted definition, even if attempts have been made to define cyber weapons [7].

The US Constitution specifically prohibits this in Article 1, Section 9: "No Bill of Attainder or ex-post facto Law shall be passed [8]."

The Second Amendment does not address the intent of the arms owner. In 1791, the right was derived from the status of a citizen in the newborn union of free colonies. The military-design, semi-automatic rifles that are protected by the Second Amendment and owned by citizens are primarily hunting rifles—in other words, a utility that could, if necessary, serve as a weapon of military conflict. There is a similarity between the gun-powder-propelled arms that are protected by the Second Amendment and dual-use hacking software, since the software is dual-use as both utility and weapon.

### V. Unusual Weapons

In the legal precedence [9], as exemplified by English v. State, 35Tex. 473, 476 (1871), Bowie-knives, slungshots, daggers, and brass knuckles with no military value do not bring Second Amendment protection to bear. In the digital realm, the unusual is far harder to prove for the government when the weaponized software is in most cases of dual use. A computer user can use a port scanner to determine which ports are responding to calls to install a printer. Thus, the port scanner could be a hacking tool, but it could also be a civilian utility software. Networking mapping software, such as the open-source software Nmap [10], is primarily a tool for network discovery. Hackers could use Nmap software to increase their knowledge of network topography, but Nmap is primarily a utility tool, not a weaponized, hacking tool. A way to weaponize the the tool is to cluster multiple Nmap achieve an ability to scan large portions of a network [11]. So the fact that clustered Nmap can be a weaponized tool and the single use of the software would give military and intelligence hackers a tool to do network discovery, even if never intended to be a military tool, render Nmap status as a military useful software. This dual use pertains not only in civilian and military realms, but also in the multiple purposes of cyberweapons as a weaponized software, and a networking utility undermines government support for framing a hacker tool as an "unusual" weapon [12]. If used by the defense establishment and the intelligence community a claim that it is an "unusual weapon" is nullified, because government through actions have evidence that it is a "usual weapon". Because hacker and discovery tools are commonly used by both the military and civilians, both communities utilize the same over-arching information technology such as wireless, sharing, data storage, Internet of Things, and communication [13],[14], the legal foundation for prohibiting civilian ownership of potentially weaponized software evaporates.

## VI. CORPORATE CYBER RIGHTS TO BEAR ARMS

In the originalist thought, militiamen were volunteers ready to defend their state and community from hostile forces, and by doing so, risked their lives in the engagement. The militia is made up of citizens of the state that have the right to bear arms. A militia risks the lives of its members to defend its community and state. Therefore, *sui genesis* a militia is mortal: a militiaman could die from combat wounds or succumb to disease and hardship as a result of combat.

The relationship between the militia, armed citizens, and lawmakers is a social contract where all parties surrender and receive. Citizens have the right to bear arms, but citizens are assumed to be reasonable—what in Roman legal tradition are described as *bonus pater familias*: the reasonable good family—a father with sound judgment and values who will provide security when needed and defend his community. In the originalist view, the militia is presumed loyal to the state that emerged after the American Revolution. The Bill of Rights was ratified 1791, less than 10 years after the end of the American Revolutionary War, which divided the colonists of the 13 colonies into two camps: American revolutionaries and loyalists to the King of England. The militia gives the state manpower to protect the people of the state and the state itself if needed, surrender their freedom partly by volunteering as militiamen and accepting a risk of loss of life, and gain a right to bear arms. Each citizen that who is given the right to bear arms does not need to be a member of the militia to execute and take favor of the right to bear arms, but the militia is recruited from the armed citizenry.

Federal US law establishes that a corporation is a US person under federal laws and executive orders. The US tax code Internal Revenue Code Section 7701(a)(30) defines a US person as "a citizen or resident of the United States (including a lawful permanent resident residing abroad who has not formally notified the United States Citizenship and Immigration Services in order to abandon that status); a domestic partnership; a domestic corporation." The National Security Agency, supported by US Presidential Executive Order 12333, states the following: "Federal law and executive order define a US person as a citizen of the United States; an alien lawfully admitted for permanent residence; an unincorporated association with a substantial number of members who are citizens of the U.S. or are aliens lawfully admitted for permanent residence; or, a corporation that is incorporated in the U.S."

The question is whether any corporation can claim Second Amendment rights to own military-grade cyber weapons as a part of the recruitment pool for the state militia and as free men with the right to bear arms.

*1) Citizens enter a social contract by bearing arms*

The argument put forward in this manuscript is that only a corporation that risks corporate death and demonstrates unquestionable loyalty to the United States could claim Second Amendment rights to own military-grade cyber arms. A corporate death is the dissolution of the corporation and liquidation of assets, often through a non-restructuring bankruptcy. If a corporation is large enough to survive and sustain operations after a failed cyber engagement, it cannot have Second Amendment rights to access military-grade cyber weapons. Therefore, any larger corporation would be a US person, but would not meet the criteria to bear arms under the Second Amendment. The larger corporation is not a part of the social contract between the framers of the Bill of Rights and, at the time, the political leadership of the United States of America and its states, as well as the citizens of those states. The large corporation could instead be seen as a rich landowner who wants to arm his subordinates and workers, creating a private army within the nation-state, and not embody a militiaman. The government could entrust larger corporations to own and utilize military-grade cyber weapons, but ownership is based on a unilateral decision by the government and is not founded on the Second Amendment.

*2) Those who bear arms are loyal to the United States of America*

Loyalty to the United States of America is a qualifier for the right to bear arms, and if not expressed at the individual level, then a corporation that forms a smaller militia would not gain the right to own cyber weapons unless its loyalty to the state is unquestionable. The Second Amendment to the US Constitution reads as follows: "A well-regulated militia, being necessary to the security of a free state, the right of the people to keep and bear arms, shall not be infringed." The militia is, according to the framers of the Bill of Rights, "necessary to the security of the free state." A defender and provider of security to the free state must be loyal to the state. Otherwise, the right to bear arms would provide a right to arm loyalists to the English Crown, and in the context of contemporary America any group that wants to overthrow the American government or support enemies of the state, as well.

In 1791, the United States had only eight years earlier emerged as a free nation liberated through a rebellion against a former colonial power that lasted seven years. The American

Revolution split families and communities. Friends became enemies as colonists chose either to join the revolution or to support continued British rule of the 13 colonies as loyalists to the Crown.

The framers of the Bill of Rights sought to protect the people from what they saw as oppression under British rule. In the Second Amendment, a prerequisite is an unquestionable loyalty to the newborn nation and its states. The right to bear arms is reserved only for those loyal to the new republic. The Bill of Rights catalogs what the rebellious colonists considered wrong with British rule and provided remedies to protect the rights of the people of the new republic from abuse, tyranny, and absence of the rule of law. The rationale for the Bill of Rights is that it provided a legal foundation for securing the freedom and liberty gained through the American revolution. As an example, the Third Amendment reads as follows: "No soldier shall, in time of peace, be quartered in any house without the consent of the owner, nor in time of war but in a manner to be prescribed by law." The Third Amendment is a safeguard against the British practice of quartering soldiers in civilian homes without providing compensation or seeking homeowners' consent.

Similarly, the Fourth Amendment states: "The right of the people to be secure in their persons, houses, papers, and effects against unreasonable searches and seizures shall not be violated, and no warrants shall issue but upon probable cause, supported by oath or affirmation, and particularly describing the place to be searched and the persons or things to be seized." The Fourth Amendment protects against the British practice of arbitrarily and without any factual support conducting searches and seizures of colonists' property and dwellings. The Fourth Amendment introduces probable cause as a threshold for government intervention.

Larger corporations fail to meet the mortal requirement in the social contract between the militiamen and the state, and also fail to meet the loyalty requirement because they operate and have ties with foreign countries; therefore, larger corporations, and especially multinational corporations, have no Second Amendment right to own or use military-grade cyber weapons.

Therefore, only smaller, non-multinational corporations that can fail and be forced to dissolve and liquidate as a result of a cyber engagement can claim the right to own and use military-grade cyber weapons under the Second Amendment.

### VII. PROHIBITION OF ZERO-DAY EXPLOITS

A zero-day exploit is a vulnerability that is unknown to anyone except its discoverer [15]. The IT industry, the computer security community, and the defense establishment are unaware of the vulnerability. After the zero-day exploit is found, a tool could be designed to take full advantage of the exploit.

This zero-day exploit tool could be "dangerous and unusual," but still is permissible to own based on the fact that the exploit the tool is targeting is unknown to the government. In any country that has submitted to the rule of law and democratic foundations, the government cannot ban and criminalize what it doesn't know or create all-encompassing penal codes for such scenarios.

A government is unaware of a zero-day exploit and until the utilization of the zero-day exploit, not only the exploit is unknown, but also the extent of the effects and damage created by its exploitation. A tool that exploits zero-day exploits could only be prohibited after execution.

### VIII. THE CASTLE DOCTRINE

The scope of this paper is cyber arms in the light of the Second Amendment. In the public discourse, the Second Amendment and the Castle Doctrine tend to part of the same discussion. The Castle Doctrine has also been used as an argument to support legalization of corporate hack back [16].

The Castle Doctrine is a common-law doctrine [17] that supports the individual's option, without prosecution or penalty, to use force in defending his residence and family. If attacked, there traditionally has been an obligation to retreat to avoid bodily harm and risk to human life. The Castle Doctrine supersedes the obligation to retreat if the attack occurs in the personal realm that the doctrine covers based on state-level precedence. In some states, the doctrine extends to an individual's vehicles and workplace. The Castle Doctrine's lineage extends to older English common law, under which a citizen was considered to have a right to peace and safety in his own home. The Second Amendment and the Castle Doctrine are separate; where the Second Amendment is the federal right to bear arms, the Castle Doctrine is state-level legislation to define the boundaries between excessive force, the obligation to retreat, and the use of lethal force to defend one's life and property. The Second Amendment right to own and bear arms does not provide a right to protect life and property using deadly or significant force. Therefore, any citizen's utilization of military grade cyber weapons owned and procured under the rights given by the Second Amendment to protect himself from physically harmful cyber attacks is contingent not only on federal legislation as exemplified by the Computer Fraud and Abuse Act (CFAA)[18], but also by state legislation. As our society merges human ability and physiology with digital mechanisms and initiates a broader use of digital-human enhancement, creating a personal attack surface for digital arms, the question of a cyber Castle Doctrine is a useful analogy from the non-digital reality.

### IX. CONCLUSION

Currently, there is no cyber Second Amendment case of the first impression on the docket. Such a case would trigger a legal development and clarification, but already a distinct legal doctrine is visible. According to the inquiry, the citizenry has the right to own and operate cyber arms under identical conditions as firearms. The current US legal doctrine states that "unusual and dangerous" arms can be prohibited. The majority of the hacking, network discovery, and information systems probing and exploiting tools are of military utility and value as enablers of cyber capacity even if in civilian hands [19]. Therefore, these tools cannot be considered unusual in the light of the Second Amendment.

The question then becomes whether these cyber arms are dangerous. The legal doctrine has developed an allowance for prohibiting dangerous weapons that would harm the public beyond an intended military use or that lack safeguards such as intent by arming and using the weapon. The labeling of a weapon as "dangerous" would require that a cyber weapon autonomously initiate attacks without any human interaction, as intent is not required when the weapon is assessed on its own merits, and this is not applicable to the vast majority of cyber weapons. An example of what could be considered dangerous could be software that autonomously attacks other information systems at the bootup of a networked computer. The software has no controlled harm creation, but instead—without any intent at the moment of engagement, nor human control—starts attacking other systems. Such software could be prohibited.

The current Second Amendment doctrine would not exclude cyber arms unless these arms are of no military value or not suitable for military use, or require no intent to be dangerous for the general population. The Second Amendment protects individual ownership of cyber arms, and to a degree corporate ownership, and the case for government intervention and prohibition of cyber arms faces significant constitutional limitations to be a realistic policy option.

ACKNOWLEDGMENT

The author thanks Robert E. Barnsby, JD, and Dr. Greg Conti for valuable input and information sharing.

JAN KALLBERG is currently an Assistant Professor of Political Science with the Department of Social Sciences, United States Military Academy at West Point, and a Research Scientist with the Army Cyber Institute at West Point. He was earlier a researcher with the Cyber Security Research and Education Institute, The University of Texas at Dallas, an Assistant Professor at Arkansas Tech University, and a part-time faculty member at George Washington University. Dr. Kallberg earned his PhD and MA from the University of Texas at Dallas and earned a JD/LL.M. from Juridicum Law School, Stockholm University. Dr. Kallberg is a certified CISSP and ISACA CISM. He has authored papers in the *Strategic Studies Quarterly*, *Joint Forces Quarterly*, *IEEE IT Professional*, *IEEE Access*, *IEEE Security and Privacy*, and *IEEE Technology and Society*.